\documentclass[twocolumn,showpacs,amsmath,amssymb]{revtex4-1}

\usepackage{amsmath}
\usepackage{amssymb}
\usepackage{graphicx}
\usepackage{color}
\usepackage{epsfig}
\usepackage{amsmath}
\usepackage{amssymb}
\renewcommand{\Im}{\mathrm{Im}\,}

\DeclareMathAlphabet{\bi}{OML}{cmm}{b}{it}

\begin{document}

\title{Magneto-optical transport properties of monolayer phosphorene}
\author{M. Tahir$^{1,\star}$, P. Vasilopoulos$^{1,\dag}$, and F. M. Peeters$^{2,\ddag}$}
\affiliation{$^{1}$Department of Physics, Concordia University, Montreal, Quebec, Canada H3G 1M8}
\affiliation{$^{2}$Departement Fysica, Universiteit Antwerpen Groenenborgerlaan 171, B-2020 Antwerpen, Belgium}

\begin{abstract}

The electronic properties of monolayer
phosphorene are  exotic due to its puckered structure and large
intrinsic direct band gap. We derive and discuss its band structure  in the presence of a perpendicular magnetic field. Further, we evaluate  the magneto-optical Hall and  longitudinal optical 
conductivities, as  functions of temperature, magnetic field, and Fermi
energy, and show that  %. Due to the structure's anisotropy t
they are strongly influenced by the
 magnetic field. The imaginary part of the former and the real part of the latter  exhibit regular 
 {\it interband} oscillations as functions of the frequency $\omega$ in the range
 $\hslash\omega\sim 1.5-2$ eV. Strong {\it intraband} responses in the latter   and week ones in the former occur at much lower frequencies. The magneto-optical response can be tuned in  the microwave-to-terahertz and % the 
 visible
frequency ranges in contrast with %\textbf{
a conventional two-dimensional electron gas or graphene in which 
the response is limited to %occurs %optical features appear in 
 the terahertz  regime. This ability to isolate carriers in an anisotropic structure may
make phosphorene a promising candidate for   new optical devices.% including thin-film polarizers, sensors, etc.
\end{abstract}

\maketitle

\section{Introduction}

Graphene possesses extraordinary
properties but its application in device fabrication is limited by the zero
band gap. Graphene-based transistors suffer from a low on-off current ratio due
to its gapless structure \cite{KS,LL}. This has lead to an intensive search for 
materials with a finite band gap including silicene \cite{JS},
germanene \cite{ME}, MoS$_{2}$, and other group-VI transition-metal
dichalcogenides \cite{QH}. Despite the fact that these materials  show a high on-off
ratio,  their carrier mobility is considerably lower than that of graphene and restricts their applications in  
electronics and optoelectronics \cite{BR}. Moreover, a very large intrinsic
direct band gap in these materials, render them unsuitable for near-infrared
optical telecommunication and mid-infrared applications. Thus the search
continues for a two-dimensional (2D) semiconducting material, with a direct band gap, high
carrier mobility, and with  the potential to form excellent contacts with known
electrode materials.

Recent developments in the experimental realization of 2D phosphorene has
attracted much interest due to its potential applications \cite{LLI,HL, XFW}. Unlike
graphene, MoS$_{2}$ and other related materials, electrons in phosphorene
are highly dispersive and delocalized along the out-of-plane direction \cite%
{CQ}. Phosphorene has a honeycomb structure of black phosphorous with large
intrinsic direct band gap of $1.52$ eV  assessed by  
tight-binding \cite{AN} and density functional theory \cite{YD,%FX,
JQ,AS}.
Compared to graphene, the puckered structure of phosphorene exhibits lower
symmetry that gives rise to in-plane anisotropic properties in momentum space.
Carrier mobilities are very high at room temperature and exhibit a  strongly
anisotropic behaviour in phosphorene-based transistors with a high on-off ratio 
\cite{HL,LLI,TL}. Due to its highly dispersive band structure, phosphorene
exhibits a high  carrier mobility and   a large optical conductivity \cite%
{JQ}. A linear dichroism has been shown in the computed absorption
spectra, in that the positions of the lowest energy absorption peaks differ strongly for the
two in-plane directions \cite{JQ}. In addition to the
traditional mid- and near-infrared emission, 
the unique anisotropic electronic and photonic properties of phosphorene may allow
for the realization of novel optical components such as polarization sensors
and anisotropic plasmonic devices \cite{TL}. Already due to its high optical
efficiency \cite{MB}, phosphorene has shown a high
potential for optical device applications \cite{NYC}.

Optical transport properties of graphene  show 
good agreement between theory and experiments \cite{VP}. Magneto-optical
properties of topological insulators \cite{WK} and other single-layer
materials, such as MoS$_{2}$ \cite{ZL} and silicene \cite{VPG}, have also been
studied. Landau levels (LLs) are formed in the presence of an external
magnetic field. Transitions between the LLs generate absorption lines in the
magneto-optical conductivity \cite{VPG} and  
were used to distinguish between the topological insulator  
and normal (band) insulator phases in
silicene in the presence of spin-orbit interaction and staggered potential 
\cite{CJ}. From a fundamental point of view, many efforts have been made to
explore different properties of phosphorene at zero magnetic field whereas
limited work has  appeared for finite magnetic field \cite{XY}. Accordingly, 
studies of magneto-optical properties are  timely %appropriate 
and expected to
increase our understanding of this material.  As will be shown, an important difference  with graphene and other 2D systems is that their  magneto-optical response occurs  in the terahertz (THz) regime whereas phosphorene's can be tuned in  the microwave-to-THz and   visible frequency ranges.  %optical features appear

In this work we study the magneto-optical transport properties of
monolayer phosphorene. In Sec. II we derive and discuss its band structure  
 in the presence of a perpendicular magnetic field.  
Further, using Kubo-type formulas in Sec. III we evaluate the optical Hall  
and  longitudinal conductivities.  We proceed with a discussion of the results  
and of  the %pertinent  
power absorption spectrum in Sec. IV. We then show briefly the 
oscillator strength of the optical transitions in Sec. V and conclude  in Sec. VI.

\section{Basic expressions}

We start with the widely used two-band model for 2D %two-dimensional 
phosphorene 
\cite{AN,XY}, in which the low-energy Hamiltonian is %given by 
\begin{equation}
%\hspace*{-0.5cm}
H=\left( 
\begin{array}{c}
\hspace{-1cm}E^{e}+(\alpha ^{\prime }\Pi _{x}^{2}+\beta \Pi _{y}^{2})/2 \quad \quad \quad 
0\\%
\ \\
 \quad  \quad  \quad 0 \quad \quad \quad \quad  
E^{h}-(\lambda ^{\prime }\Pi _{x}^{2}+\eta \Pi _{y}^{2})/2%
\end{array}%
\right),  \label{1}
\end{equation}
where $\alpha ^{\prime }=\alpha +\gamma ^{2}/E_{g}$, $\lambda ^{\prime
}=\lambda +\gamma ^{2}/E_{g}$, $\gamma =8.5\ast 10^{5}$ m/s$,\alpha
=1/m_{ex}=1/1.47m_{e}$, $\beta =1/m_{ey}=1/ 0.83m_{e}$, 
$\lambda =1/m_{hx}=1/10.66m_{e}$,  $\eta =1/m_{hx}=1/
1.12m_{e}$, and $E_{g}=E^{e}-E^{h}=1.52$ eV. 
The minimum of the conduction band  occurs at $E^{e}=0.34$ eV and the maximum of the valence band at $E^{h}=-1.18$ eV.  Further,  $\mathbf{\Pi =p}+e\mathbf{A}$ is the 2D
canonical momentum with vector potential $\mathbf{A}$.  Using the
Landau gauge $\mathbf{A}= (0, Bx, 0)$ and diagonalizing the
Hamiltonian  (1), we obtain the eigenvalues   
\begin{equation}
E_{n}^{s}=E^{s}+s(n+1/2)\hslash \omega _{c}^{s}\text{, }n=0,1,2,3,...
\label{2}
\end{equation}
where $s=+1( -1)$ for electrons (holes), $E^{s}=E^{e/h}$, $\omega _{c}^{s}=\omega
_{c}^{e/h}$ with $\omega _{c}^{e}=eB/\sqrt{m_{ex}^{\prime }m_{ey}}%
=2.696\omega _{c}$, $\omega _{c}^{h}=eB/\sqrt{m_{hx}^{\prime }m_{hy}}%
=2.2076\omega _{c}$, $m_{ex}^{\prime }=1/\alpha ^{\prime }$, $%
m_{hx}^{\prime }=1/\lambda ^{\prime }$, and $\omega _{c}=eB/m_{e}$. It is
interesting to note that unlike the anisotropic zero magnetic field
dispersion, the LL spectrum is independent of the in-plane wave vectors. The
corresponding normalized eigenfunctions are  
\begin{equation}
\hspace*{-0.33cm}\Psi _{n,k_{y}}({\bf r}%x,y
)=\frac{e^{ik_{y}.y}}{\sqrt{L_{y}}}\Big( 
\begin{array}{c}
\phi _{n}(u^{e}) \\ 
\phi _{n}(u^{h})%
\end{array}%
\Big),  \label{3}
\end{equation}
where %we set 
$u^{s}=\xi^{s}(x-x_{0})/l$ and   
$\xi^{s}=\sqrt{m_{sx}^{\prime }\omega
_{c}^{s}/\hslash }$; $l=\sqrt{\hslash /eB}$ is the magnetic length and $\phi _{n}(u)$ are the harmonic oscillator functions.  If we use the Landau gauge 
$A=(-By,0,0)$, the eigenvalues are again given by Eq. (2) and %. The corresponding 
the eigenfunctions are obtained 
from Eq. (3) by replacing $x$ with $y$ wherever they appear.

The density of states (DOS) is given by
\begin{equation}
D(E )=\frac{1}{S_{0}}\sum_{n,s,k_{y}}\delta (E
-E_{n}^{s}),  \label{4}
\end{equation}
where $S_{0}$ is the area of the system. The sum over $k_{y}$ in Eq. (4) is
evaluated using the prescription ($k_{0}=L_{x}/2l^{2}$) $\sum_{k_{y}}%
\rightarrow (L_{y}/2\pi)g_{s}\int_{-k_{0}}^{k_{0}}dk_{y}=(S_{0}/
D_{0})g_{s}$, where $D_{0}=2\pi l^{2}$ and $g_{s}=2$ is the spin degeneracy.  
The Fermi energy $E_F$ is determined from the electron
concentration $n_{c}$,  
\begin{equation}
n_{c}=\int_{-\infty }^{\infty }D(E )f(E )dE
 =g_{s}/D_{0}\sum_{n,s}f(E_{n}^{s}),  \label{5}
\end{equation}%
where the Fermi-Dirac distribution function is written as $%
f(E_{n}^{s})=(1+\exp [\beta (E_{n}^{s}-E)])^{-1}$ with $\beta
=1/k_{B}T$. In Fig. 1 (top panel) the magenta solid curve shows $E_F$  obtained
numerically from Eq. (5) as a function of $B$  for a realistic value  of the electron density \cite{XY}, $%
n_{c}=1\times 10^{16}$ m$^{-2}$,  together with the LLs obtained from Eq. (2). 

Assuming a Gaussian broadening  of the LLs,  Eq. (4) is rewritten 
 as $D(E)=(g_{s}/D_{c})\sum_{n,s}\exp \big[ -(E
-E_{n}^{s})^{2}/2\Gamma ^{2}\big]$,
 where $D_{c}=D_{0}\Gamma \sqrt{2\pi }$ and 
$\Gamma \propto \sqrt{B}$ is the width of the
Gaussian distribution \cite{YZTA}. 
The dimensionless DOS at $E=E_F$, $D(E_F)\equiv D(B)$, is shown in Fig. 1 (bottom panel) as a
function of the magnetic field, for $\Gamma =0.2\sqrt{B}$ meV (black solid curve) and $\Gamma =0.1\sqrt{B}$ meV (red dotted curve).  For weak  
 fields $B$ the level broadening is important 
whereas for  large fields it %this effect 
may become smaller due to the $\sqrt{B}$ dependence
%\textbf{
since the distance between the LLs increases linearly with $B$.

\begin{figure}[ht]
\includegraphics[width=0.8\columnwidth,height=0.5\columnwidth]{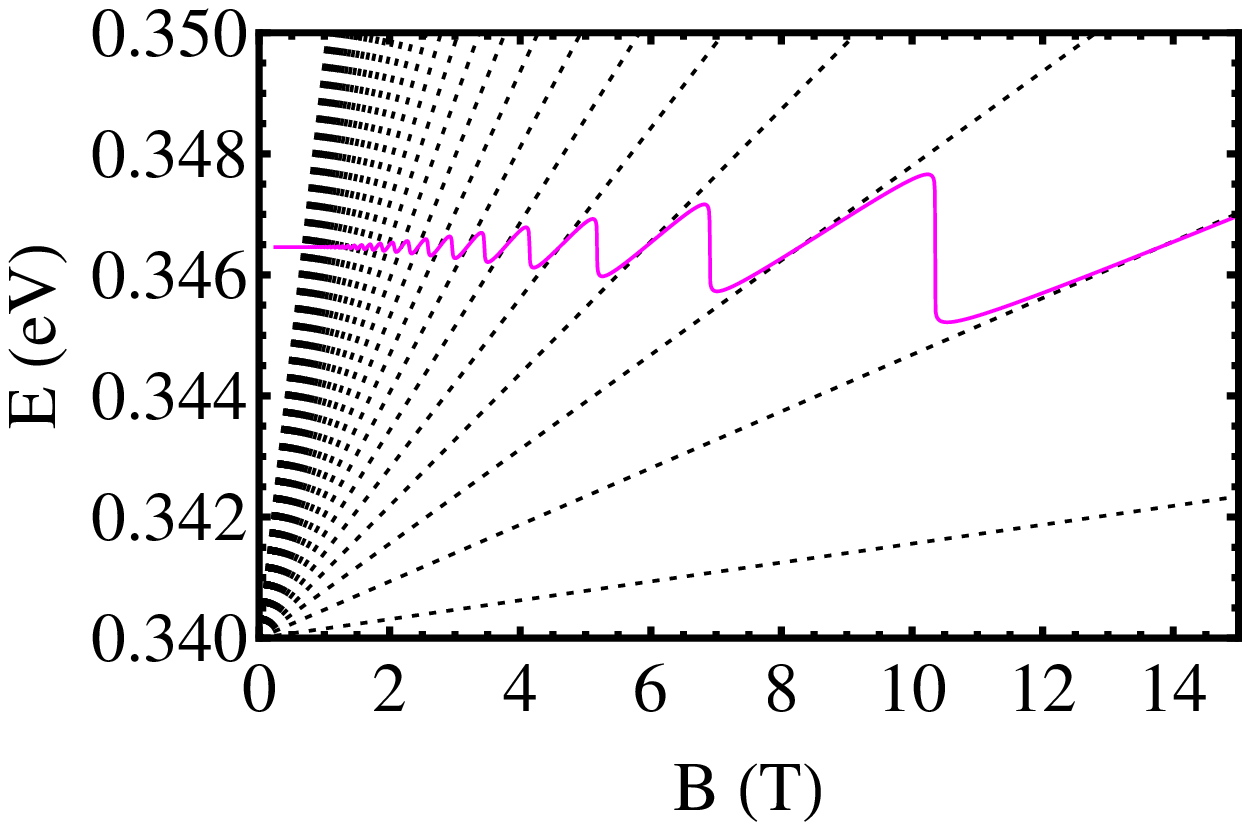}L
\hspace*{0.6cm}
\includegraphics[width=0.72\columnwidth,height=0.5\columnwidth]{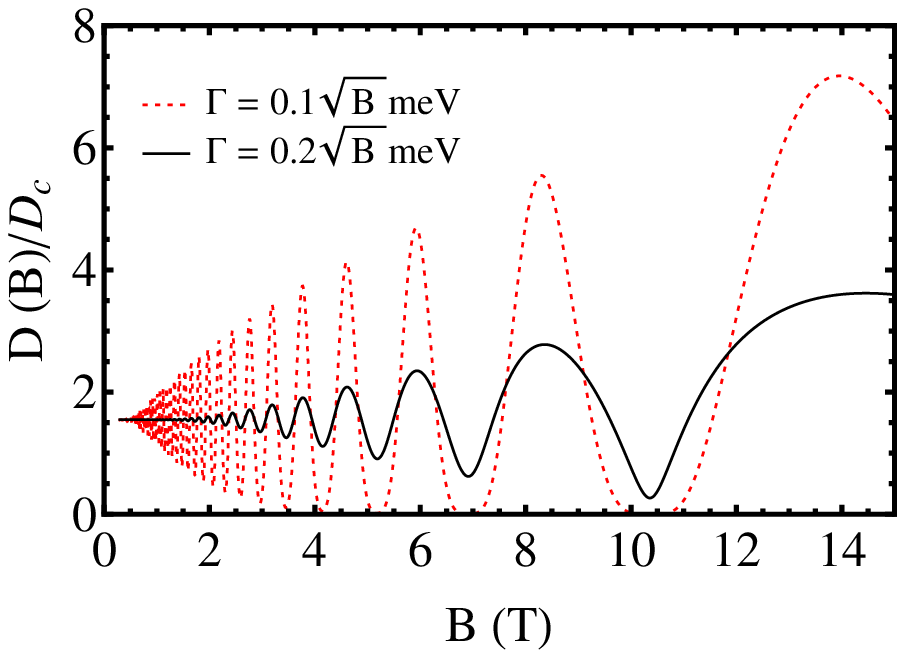}
\vspace*{-0.35cm}
\caption{Top panel: Fermi energy (solid magenta) as a function of the magnetic field for $T=1$ K and density  $n_{c}=1\times 10^{16}$ m$^{-2}$. The dashed curves are the LLs  
evaluated from Eq. (2). Bottom panel: Dimensionless density of states as a function of the magnetic field $B$ for two different values of the LL width. }%, as indicated.}
\end{figure}

\section{ Linear response  conductivities}  

We consider a many-body system described by the Hamiltonian $\hat{H}=\hat{H}%
_{0}+\hat{H}_{I}-\mathbf{\hat{R}}\cdot \mathbf{\hat{F}}(t)$, where $\hat{H}%
_{0}$ is the unperturbed part, $\hat{H}_{I}$ is a binary-type interaction
(e.g., between electrons and impurities or phonons), and $-\mathbf{\hat{R}}%
\cdot \mathbf{\hat{F}}(t)$ is the interaction of the system with the
external field $F(t)$ \cite{MC}. For conductivity problems we have $\mathbf{%
\hat{F}}(t)=e\mathbf{\hat{E}}(t)$, where $\mathbf{\hat{E}}(t)$ is the
electric field, $e$ the electron charge, $\mathbf{\hat{R}}=\sum_{i}\mathbf{%
\hat{r}}_{i}$, and $\mathbf{\hat{r}}_{i}$ is the position operator of
electron $i$. In the representation in which $\hat{H}_{0}$ is diagonal the
many-body density operator $\hat{\rho}=\hat{\rho}^{d}+\hat{\rho}^{nd}$ has a
diagonal part $\hat{\rho}^{d}$ and a nondiagonal part $\hat{\rho}^{nd}$. For
weak electric fields and weak scattering potentials, for which the first
Born approximation applies, the conductivity tensor has a diagonal part $%
\sigma _{\mu \nu }^{d}$ and a nondiagonal part $\sigma _{\mu \nu }^{nd}$
part, $\sigma _{\mu \nu }=\sigma _{\mu \nu }^{d}+\sigma _{\mu \nu }^{nd}$, $%
\mu ,\nu =x,y$.

In general we have two kinds of currents, diffusive and hopping, but usually
only one of them is present. In the present problem, due to the magnetic field
we have only the hopping current since the diffusive one vanishes due to
$v_{\mu \zeta }=0$, see Eq. (6). For elastic scattering the diffusive 
part of $\sigma _{\mu \nu }^{d}$ is given by  
\begin{equation}
\sigma _{\mu \nu }^{dif}(\omega )=\frac{\beta e^{2}}{S_{0}}\sum_{\zeta
}f_{\zeta }(1-f_{\zeta })\frac{v_{\nu \zeta }\,v_{\mu \zeta }\,\tau _{\zeta }
}{1+i\omega \tau _{\zeta}},  \label{6}
\end{equation}%
where $S_{0}=L_{x}L_{y}$, $\tau _{\zeta }$ is the  relaxation time, $ 
\omega $ the frequency, and $v_{\mu \zeta }$ the diagonal matrix elements of
the velocity operator. Further, $f_{\zeta }=[1+\exp {\beta (E_{\zeta }-E_{F})%
}]^{-1}$ is the Fermi-Dirac distribution function with $\beta =1/k_{B}T$, $T$
the temperature and $E_{F}$ the Fermi level. 

The collisional part
$\sigma _{\mu \nu }^{col}( \omega )$ of $\sigma _{\mu \nu }^{d}$ is much smaller than  
the nondiagonal part  $\sigma _{\mu \nu }^{nd}(\omega )$ 
and we neglect  it. 
As for $\sigma _{\mu \nu }^{nd}(\omega )$ given in Ref. \cite{MC}, one can use the
identity $f_{\zeta }(1-f_{\zeta ^{\prime }})[1-\exp {\beta (E_{\zeta
}-E_{\zeta ^{\prime }})}]=f_{\zeta }-f_{\zeta ^{\prime }}$ and cast \cite{VPC} its
original form   in the more familiar one 
\begin{equation}
\hspace*{-0.28cm}\sigma _{\mu \nu }^{nd}(\omega )=\frac{i\hbar e^{2}}{S_{0}}\sum_{\zeta \neq
\zeta ^{\prime }}\frac{(f_{\zeta }-f_{\zeta ^{\prime }})\,v_{\nu \zeta \zeta
^{\prime }}\,v_{\mu \zeta ^{\prime }\zeta }}{(E_{\zeta }-E_{\zeta ^{\prime
}})(E_{\zeta }-E_{\zeta ^{\prime }}+\hbar \omega +i\Gamma )}\,,  \label{7}
\end{equation}
\noindent where the sum runs over all quantum numbers $\left\vert \zeta
\right\rangle \equiv \left\vert n,s,k_{y}\right\rangle $ and $\left\vert
\zeta ^{\prime }\right\rangle \equiv \left\vert n^{\prime },s^{\prime
},k_{y}^{\prime }\right\rangle $ provided $\zeta \neq \zeta ^{\prime }$. The
infinitesimal quantity $\epsilon $ in the original form has been
replaced by $\Gamma$ to account for the broadening of the energy
levels. Equation (7) has been recently applied to phosphorene in the absence of a magnetic field  in Ref. \onlinecite{MB}(b). The evaluation of  $v_{\nu \zeta \zeta ^{\prime }}$ and $v_{\mu \zeta \zeta
^{\prime }}$ is outlined in the  Appendix. Using Eqs. (1) and (3) %, with $s=s'$ 
we obtain
\begin{equation}
\hspace*{-0.45cm}
v_{x,n,n^{\prime}}^{s,s}= i v_{x}^s 
\left( \sqrt{n}\delta _{n-1,n^{\prime }}-\sqrt{n+1}%
\delta _{n+1,n^{\prime }}\right) \delta_{k_{y},k_{y}^{\prime}}, \label{8}
\end{equation}
\begin{equation}%{eqnarray}
\hspace*{-0.44cm}
v_{y,n^{\prime},n}^{s,s} = v_{y}^s
\left( \sqrt{n+1}\delta _{n^{\prime },n+1}+\sqrt{n}\delta _{n^{\prime
},n-1}\right)\delta_{k_{y},k_{y}^{\prime}} ,  \label{9} 
%  \notag
\end{equation}%{eqnarray}
where $v_{x}^s=\omega _{c}^{s}/\sqrt{2}\xi^{s}$ 
and $v_{y}^s=(m_{sx}^{\prime }\omega _{c}^{s}/\sqrt{2}\xi^{s}) (\beta/ \eta) $. The results for 
$s\neq s^{\prime}$  are given in the appendix.
Since $\left\vert \zeta \right\rangle \equiv \left\vert
n,s,k_{y}\right\rangle $, there will be one summation over $k_{y}$\ which,
with periodic boundary conditions for $k_{y}$, gives the factor $S_{0}/2\pi
l^{2}$. Substituting Eq. (8) into Eq. (7),  summing  over $n'$, and setting $\sigma _{0}=-g_{s}\hbar e^{2}/2\pi l^{2}$, we obtain the longitudinal  nondiagonal conductivity as 
\begin{eqnarray} 
\nonumber
\sigma _{xx}^{nd}(\omega )&=&i\sigma _{0}\sum_{s,s^{\prime
},n=1}\frac{n\big[f_{n}^{s}-f_{n-1}^{s^{\prime }}\big](A_{xx}^{s,s^{\prime}})^2 
}{%
I_{n,n-1}^{s,s'}  (I_{n,n-1}^{s,s'}+\hbar \omega +i\Gamma )}
\end{eqnarray} 
\begin{eqnarray} 
\hspace*{0.99cm}+i\sigma _{0}\sum_{s,s^{\prime
},n=0}\frac{(n+1)\big[f_{n}^{s}-f_{n+1}^{s^{\prime }}\big](A_{xx}^{s,s^{\prime}})^2 }{
I_{n,n+1}^{s,s'} (I_{n,n+1}^{s,s'}+\hbar \omega +i\Gamma )}, \label{10} 
\end{eqnarray} 
where $A_{xx}^{s,s}=v_{x}^s$, $A_{xx}^{s,s^{\prime}}=
v_{x}^e-v_{x}^h$, and $I_{n,n\pm 1}^{s,s'}=E_{n}^{s}-E_{n\pm 1}^{s^{\prime}}$. After making the changes $n-1 \to m \to n$ in the first sum, we  combine the two sums and obtain 
\begin{eqnarray} 
\nonumber
\hspace*{-0.5cm}
\sigma _{xx}^{nd}(\omega )&=&i\sigma _{0}\sum_{s,s^{\prime
},n=0}(n+1) \Big[\frac{\big[f_{n+1}^{s}-f_{n}^{s^{\prime }}\big](A_{xx}^{s,s^{\prime}})^2
 }{ I_{n+1,n }^{s,s'} (I_{n+1,n}^{s,s'}+\hbar \omega +i\Gamma )}
\end{eqnarray}% 
\begin{eqnarray}% 
 &&\hspace*{1.99cm}+\frac{\big[f_{n}^{s}-f_{n+1}^{s^{\prime }}\big](A_{xx}^{s,s^{\prime}})^2
 }{ I_{n,n+ 1}^{s,s'} 
(I_{n,n+ 1}^{s,s'}+\hbar \omega +i\Gamma )}\Big]. \label{11} 
\end{eqnarray} 

In the limit $\Gamma\to 0,\,\omega\to 0$ and $s=s^{\prime}$ Eq. (11) yields zero.
The  matrix elements of the velocity operators are nonzero only for  
$%
n^{\prime }=n\pm 1$. Regarding the sums over $s,s'$ for convenience we write $\sum_{s,s^{\prime
}}=\sum_{+,+}+\sum_{-,-}+\sum_{+,-}+\sum_{-,+}$. Here the subscript $+/-$
denotes the conduction/valence band. After performing the summation over $%
s$ and $s^{\prime }$ we obtain the absorption spectrum of the real part of 
$\sigma _{xx}^{nd}(\omega )$ in the form
%the nondiagonal longitudinal optical conductivity%
%
\begin{eqnarray} 
\nonumber
&&\Re\sigma _{xx}^{nd}(\omega )=\pi\sigma _{0}
\sum_{n=0}^\infty (n+1) \\*
&&\hspace*{-0.7cm}\times\left[  
\begin{array}{c}
-\frac{[f_{n}^{+}-f_{n+1}^{+}\,] \delta(\hbar \omega_{c}^{e}-\hbar \omega )}{\hbar \omega_{c}^{e}/(A_{xx}^{+,+})^{2}}

+\frac{[f_{n}^{-}-f_{n+1}^{-}] \delta(\hbar \omega_{c}^{h}+\hbar \omega
)}{\hbar \omega_{c}^{h}/(A_{xx}^{-,-})^{2}}\\
\\
-\frac{[f_{n+1}^{-}-f_{n}^{+}] \delta(I_{n,n+1}^{+,-}-\hbar \omega )
}{I_{n,n+1}^{+,-}/(A_{xx}^{+,-})^{2}}

+\frac{[f_{n}^{-}-f_{n+1}^{+}] \delta(I_{n,n+1}^{-,+}+\hbar \omega
)}{I_{n,n+1}^{-,+}/(A_{xx}^{-,+})^{2}}
\end{array}%
\right]. \,\,  \label{12}
\end{eqnarray}% 
Here $I_{n,n+1}^{s,s}=E_{n}^{s}-E_{n+1}^{s}$ and $\pi\delta(x)=\Gamma/(x^2+\Gamma^2)$.
Using Eqs. (8) and (9) into Eq. (7), and carrying out the sum over $n'$ similar to Eq. (10) and then making the changes $n-1 \to m \to n$ in the first sum, we obtain the optical Hall conductivity as 
\begin{eqnarray}% 
\nonumber
\hspace*{-0.5cm}
\sigma _{xy}^{nd}(\omega )&=&\sigma _{0}\sum_{s,s^{\prime
},n=0}(n+1) \Big[\frac{\big[f_{n+1}^{s}-f_{n}^{s^{\prime }}\big]A_{xy}^{s,s^{\prime}}
 }{ I_{n+1,n }^{s,s'} (I_{n+1,n}^{s,s'}+\hbar \omega +i\Gamma )} %\\*
\end{eqnarray}%{ 
\begin{eqnarray}% 
 &&\hspace*{1.99cm}-\frac{\big[f_{n}^{s}-f_{n+1}^{s^{\prime }}\big]A_{xy}^{s,s^{\prime}} 
}{I_{n,n+ 1}^{s,s'}  (I_{n,n+ 1}^{s,s'}+\hbar \omega +i\Gamma )} \Big], \label{13} 
\end{eqnarray}% 
where $A_{xy}^{s,s}=(v_{x}^s)^{2}m_{sx}^{\prime}\beta$ and $A_{xy}^{s,s^{\prime}}=
(v_x^e -v_x^h) *(\beta m_{ex}^{\prime}v_x^e -\eta m_{hx}^{\prime}v_x^h$).  In
the $\omega=\Gamma=0$ and $s=s^{\prime}=e$ limit  Eq. (13)
yields the  quantized  Hall conductivity  of  a  2DEG \cite{VPC}
\begin{equation}
\sigma _{xy}^{nd}(0)=g_{s}\frac{e^{2}}{2h}\sum_{%s=\pm,
n=0}^\infty(n+1)\big[f_{n}-f_{n+1}\big].   \label{14}
\end{equation}
Also, apart from the absence of valley degeneracy  and the appearance of $(n+1)$ instead of $(n+1/2)$, the dc limit of Eq. (13) gives results similar to those of graphene \cite{PKPV}.

Following the  procedure adopted for
the component $\sigma _{xx}^{nd}$ and using Eq. (10) we obtain  
\begin{eqnarray}% 
\nonumber
&& \Im\sigma _{xy}^{nd}(\omega )=\pi\sigma _{0}
\sum_{n}
(n+1)\\* 
&&\hspace*{-.85cm}\times\left[ 
%\nonumber
\begin{array}{c}
-\frac{[f_{n}^{+}-f_{n+1}^{+}] \,\delta(\hbar \omega_{c}^{e}-\hbar \omega )}{\hbar \omega_{c}^{e}/A_{xy}^{+,+}}%\\
%\\
 -\frac{[f_{n}^{-}-f_{n+1}^{-}]\,\delta(\hbar \omega_{c}^{h}-\hbar \omega )}{\hbar \omega_{c}^{h}/A_{xy}^{-,-} }\\
 \\ 
+\frac{[f_{n+1}^{-}-f_{n}^{+}] \,\delta(I_{n,n+1}^{+,-}-\hbar \omega )}{I_{n,n+1}^{+,-}/A_{xy}^{+,-}}
%\\
+
\frac{[f_{n}^{-}-f_{n+1}^{+}]\,\delta(I_{n,n+1}^{-,+}+\hbar \omega )}{I_{n,n+1}^{-,+}/A_{xy}^{-,+}}
\end{array}%
\right]. \,\,  \label{15}
\end{eqnarray}%  

\section{Discussion of Results}

The energies of the positive branch levels in Eq. (2) are different from those
 of the negative branch due to the difference in the $E^{s}$ values %of the band gap 
 and
cyclotron frequency in each band (electron/hole). Due to $\hslash \omega _{c}^{s}<<E^{s}$, the {\it intraband} and {\it interband} optical transitions  belong to two widely
separated regimes: the  former is in the microwave-to-THz range and the latter in
the visible frequency range. 
We will first consider the latter that involve 
transitions between neighbouring LLs ($n^{\prime }=n\pm 1$) and $s\neq s'$. 
In all results shown below the common parameter are 
%\textbf{For further discussion, we fixed the 
temperature T = 10 K and  level
broadening $\Gamma =0.2\sqrt{B}$ meV.
%, and strong field $B=10$ T so that well-resolved  LLs  are formed.

\begin{figure}[ht]
\includegraphics[width=0.9\columnwidth,height=0.85\columnwidth]{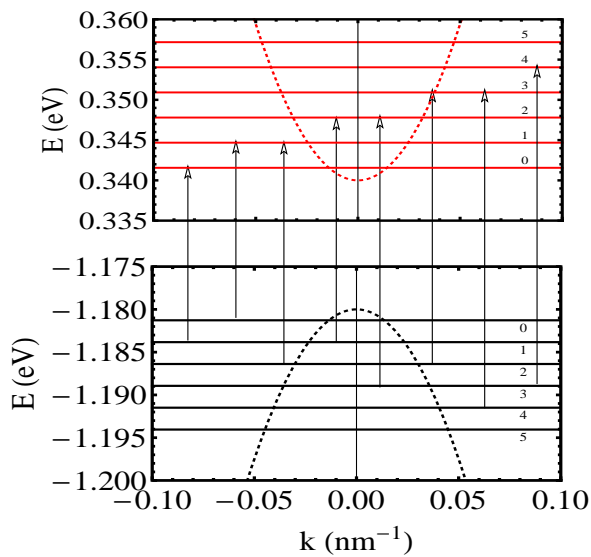}
\vspace*{-0.3cm}
\caption{Band structure at a fixed magnetic field $B=10$ Tesla for conduction (red) and valence band (black). 
The dashed curves show the $B=0$ spectrum of Eq. (1):  it is symmetric around the $\Gamma$ point and consistent with Fig. 1 of Ref. \cite{XY} (a).  The arrows indicate possible {\it interband} transitions. }
\end{figure}

Unlike
graphene \cite{VP}, silicene \cite{VPG,CJ}, and topological insulators \cite{WK}, 
the large intrinsic band gap and the lack of perfect symmetry between the positive and
negative branches of the phosphorene spectrum, shown in Fig. 2 for $B=0$ (dashed curve) and 
 $B\neq 0$ (straight solid lines), have important implications for the peaks seen in  the real  
part of $\sigma _{xx}^{nd}(\omega )$ and the imaginary parts of $\sigma _{xy}^{nd}(\omega )$.
In Fig. 3 we plot the former  as a function of the frequency. 
We consider a rather strong field $B=10$ T so that well-resolved  LLs  are formed.
The value $E_{F}=0.343$ eV is 
between the $n=0$ and $n=1$ LLs, whereas the value $E_{F}=0.356$ eV is
between the $n=4$ and $n=5$ LLs.
%The missing of first peak
 In the latter case the peaks for $n\leq 3$ are Pauli blocked and no longer possible.
We notice the equally spaced absorption peaks.  The optical selection rules
allow $n$ to change only by 1. In addition one needs to go from occupied to
unoccupied states through the absorption of a photon.  
%For fixed $E_{F}$ t
The first  peak occurring at  $\hslash \omega$ = 1.526 eV represents  transitions 
involving the $n=0$ LL. In fact, it is the sum of the absorption peaks of two  transitions
involving the energy differences $E_{1}^{+}-E_{0}^{-}$ and $E_{0}^{+}-E_{1}^{-}$ and is 
described by the  last two terms in Eq. (12). 
%\textbf{
This can also be understood from the spectrum shown in Fig. 2.
A similar explanation holds for the peaks 
%occurring 
at $\hslash \omega$ = 1.532 eV,  $\hslash \omega$ = 1.537 eV, etc. %and so on.
 The peak spacing is proportional to $B$ and  experimentally
one should observe such well-spaced peaks even for modest fields $B$. 
In contrast to phosphorene, in graphene  and other 2D systems, the spectral weight of the
interband peaks is %continuously 
redistributed into %the 
intraband peaks \cite{CJ,PEC}. This
shows how the conductivity changes as $E_{F}$ moves through the
LLs.
\begin{figure}[ht]
\includegraphics[width=0.4\textwidth,clip]{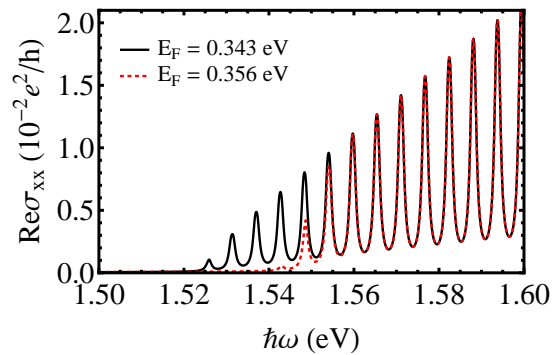}
\vspace*{-0.3cm}
\caption{Real part of the longitudinal optical conductivity as a function
of the photon energy for a field $B=10$ Tesla.}  
\end{figure}

\begin{figure}[ht]
\vspace*{-0.4cm}
\includegraphics[width=0.4\textwidth,clip]{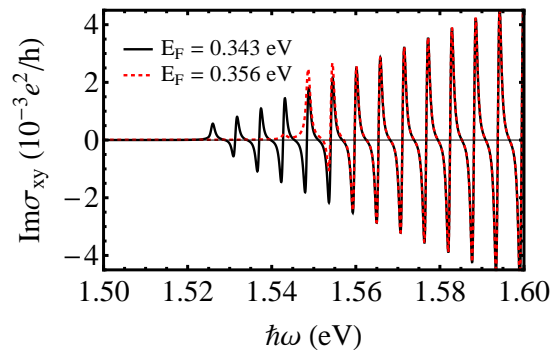}
\vspace*{-0.3cm}
\caption{Imaginary part of the optical Hall conductivity versus  photon energy
for a field $B=10$ Tesla.}
% The parameters are the same as in Fig. 3.} 
 \end{figure}

\begin{figure}[bt]
\hspace*{-0.5cm}
\includegraphics[width=0.8\columnwidth,height=0.5\columnwidth]{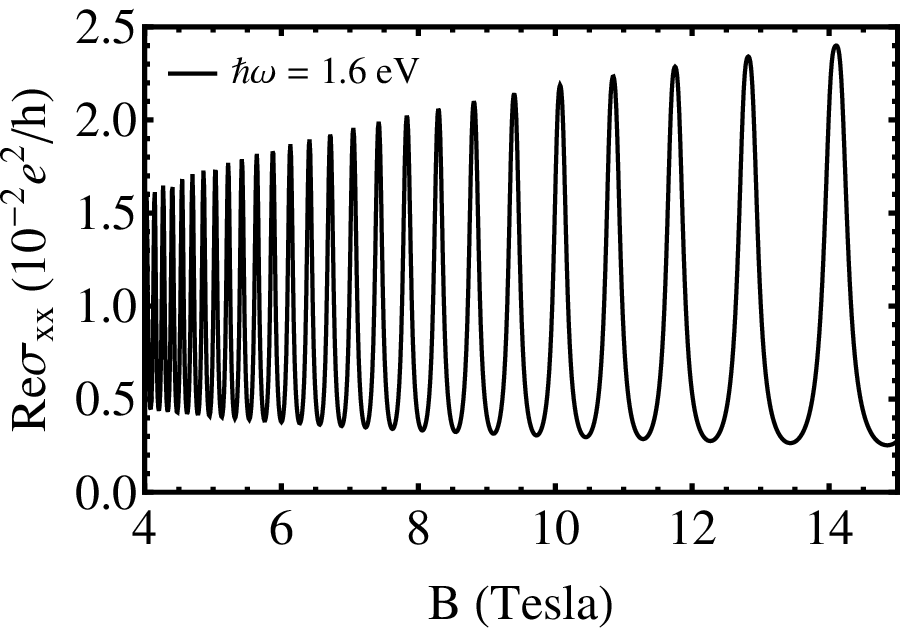}
\hspace*{-1cm}
\includegraphics[width=0.85\columnwidth,height=0.5\columnwidth]{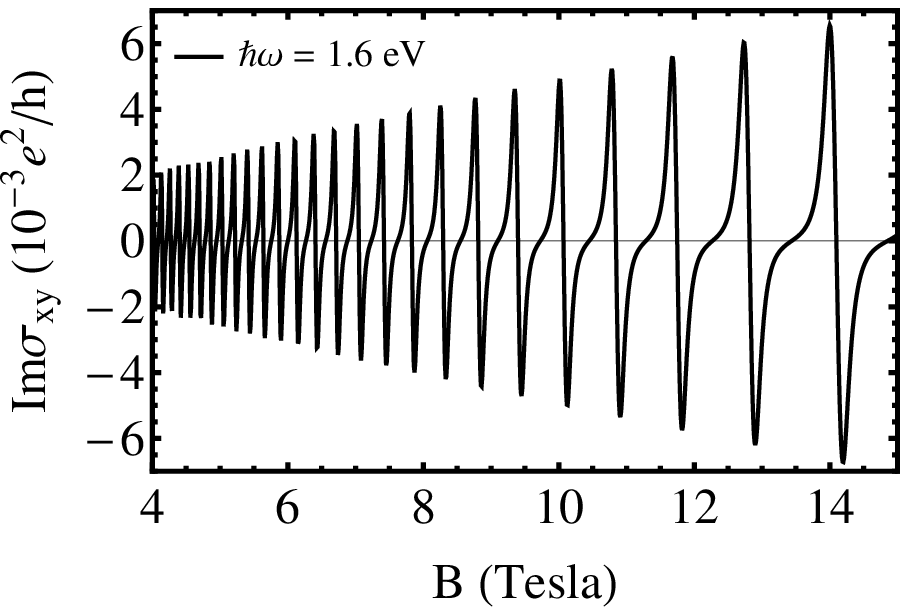}
\vspace*{-0.3cm}
\caption{Real part of the longitudinal optical conductivity (top panel) and 
imaginary part of the optical Hall conductivity (lower panel) versus  field $B$
for  photon energy $\hslash\omega=1.6$ eV.}  
\end{figure}

%The difference  between the distances $E_F-E^e_n$  and $E_F-E^h_n $, %eV from the Fermi level,
%with  $E^e_n$ and $E^h_n$ the electron and hole spectra, %given by Eq. (2), 
%enters the Fermi functions in the results, see the last two terms in 
Inspection of Eqs. (12) and (15) shows that the Hall  and longitudinal conductivities are 
different due to the factors $A_{xy}^{ss'}$ and $A_{xx}^{ss'}$ that reflect the difference between Eqs. (8) and (9).
 % than on the %are not symmetrically located with respect to the centre  of the gap. %and the differences . % or the Fermi leveand so is the case of the corresponding LLs.
%This difference, %"asymmetry", %between the electron and hole spectra,  shown in Fig. 2, 
%it has a stronger effect on  the  Hall conductivity. % than on the longitudinal one.
In Fig. 4 one  first sees a positive peak and then a decrease (dip).  
This represents   {\it interband} transitions involving the $n=0$ LL  similar to Fig. 3 and   
the energy differences 
$E_{1}^{+}-E_{0}^{-}$ and $E_{0}^{+}-E_{1}^{-}$. It can be understood as  the sum of the last two terms in Eq. (15). For the pure Dirac case these two peaks would
occur at the same energy and hence would cancel out perfectly due to the symmetry of the spectrum. Only the first peak would remain in the Hall
conductivity and all higher peaks  would cancel out \cite{CJ,PEC}. 
%\textbf{The magnitude of the longitudinal optical conductivity is higher than the optical Hall conductivity due to different values of the velocity components along x-and y-axis.}

\begin{figure}[ht]
\vspace*{-0.2cm}
\includegraphics[width=0.8\columnwidth,height=0.5\columnwidth]{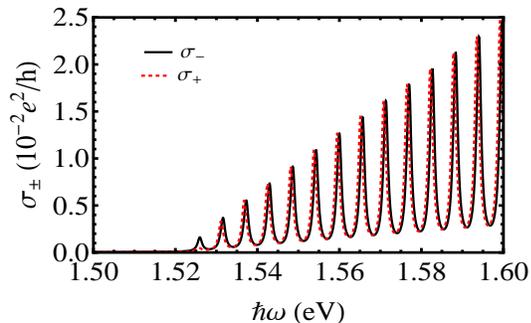}
\vspace*{-0.4cm}
\caption{Real part of the right-polarized optical conductivity 
$\sigma _{+ }(\omega )$ and of the left-polarized one 
$\sigma _{- }(\omega )$  versus  photon energy for $E_F=0.343$ eV and  field $B=10$ Tesla.}
\end{figure}

In Fig. 5 
we show the real part of the longitudinal  conductivity (top panel) and 
the imaginary one of the  Hall conductivity (lower panel) versus the field $B$
for   $\hslash\omega=1.6$ eV. % and $T=10$ K. 
Due to the interband transitions from occupied to unoccupied states 
through the absorption of the photon, the oscillation patterns are similar 
to those shown in Figs. 3 and 4.

The peak structure just described above for $\Re\sigma
_{xx}^{nd}(\omega )$ and $\Im\sigma _{xy}^{nd}(\omega )$  importantly
affects their behaviour %of the optical conductivities 
for  right ($+$) and  left  ($-$) polarized light. For real experiments that probe the (circular) polarization of resonant light, as in the case of the Kerr and Faraday effects,
one evaluates the quantity $\sigma _{\pm }(\omega )$ given by 
\begin{equation}
\sigma _{\pm }(\omega )=\Re\sigma _{xx}^{nd}(\omega )\mp \Im\sigma
_{xy}^{nd}(\omega),  \label{16}
\end{equation}
with the upper (lower) signs corresponding to right (left) polarization  \cite{VP,WK}.
 In Fig. 6 we show $\sigma
_{-}(\omega )$ (solid black curve) and  $\sigma _{+}(\omega )$ (dotted red curve)  as functions of the frequency using the parameters of Figs. 3 and 4. As seen, there is a direct correspondence 
between these results and those 
of Figs. 3 and 4.  The
peaks in $\sigma
_{+}(\omega )$   are shifted a bit (downward) in energy relative to those in $\sigma _{-}(\omega )$.
This difference also shows up in  the power absorption 
spectrum given by
\begin{equation}
\hspace*{-0.15cm}
P(\omega )=%\frac{E^{2}}{2}
(E/2)\Big[\sigma _{xx}(\omega )+\sigma _{yy}(\omega )-i\sigma _{yx}(\omega )+i\sigma _{xy}(\omega )\Big]. \label{17} 
\end{equation}%

We remind that $\sigma _{\mu \nu }=\sigma _{\mu \nu }^{d}+\sigma _{\mu \nu }^{nd}=\sigma _{\mu \nu }^{nd}$ since the component $\sigma _{\mu \mu }^{d}, \mu=x,y,$ vanishes.
The component  $\sigma _{yy}^{nd}(\omega )$ is given by $\sigma _{xx}^{nd}(\omega )$ with 
$A_{xx}^{s,s^{\prime}}$ replaced by $A_{yy}^{s,s^{\prime}}$, and
$\Im\sigma _{xy}^{nd}(\omega )=-\Im\sigma _{yx}^{nd}(\omega )$.  The spectrum $P(\omega )$ is shown in Fig. 7 as a function of the photon energy for two values of $E_F$.
Given that $\Im\sigma _{xy}^{nd}(\omega )$ is much smaller than $\Re\sigma _{xx}^{nd}(\omega )$, cf. Figs. 3 and 4, the peaks in it are  essentially the same as  those in the  longitudinal conductivity. The absence of the $n\leq 3$ peaks for $E_{F}=0.356$ is due to Pauli blocking  and consistent with Figs. 3 and 4.
\begin{figure}[ht]
\vspace*{-0.15cm}
\includegraphics[width=0.8\columnwidth,height=0.5\columnwidth]{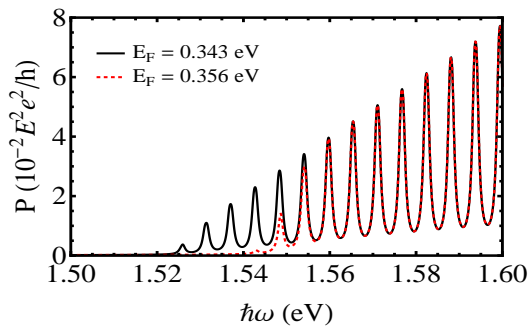}
\vspace*{-0.3cm}
\caption{Power spectrum vs photon energy for an electric field $E=8$ V/nm,
for two values of $E_F$ and  field $B=10$ Tesla.}
\end{figure}

Now we consider {\it intraband} transitions between the $n$th and ($n+1$)th LLs, for $E_{F}>0$, which  involve an  energy range 
 much smaller than $E_{F}$.  This involves large values of $n$ and is known as the semiclassical limit of the magneto-optical conductivity in which  $E_{F}$ is much larger than $\hbar\omega_c$. Let us assume that $E_{F}\approx E_{n}^{+}$ lies  
 between the $n$th and ($n+1$)th LLs. The pertinent energy difference  is   $ 
 E_{n}^{+}-E_{n+1}^{+}=-\hslash \omega _{c}^{e}$. For such transitions we   obtain
\begin{equation}
\Re\sigma _{xx}^{nd}(\omega )=\pi\sigma _{0}
\sum_{n}(n+1)\frac{[f_{n+1}^{+}-f_{n}^{+}]}{\hslash \omega _{c}^{e}/(A_{xx}^{+,+})^{2}}
\, \delta(\hslash \omega _{c}^{e}-\hbar \omega
).  \label{18}
\end{equation}
%
%The real part of $\sigma _{xx}^{nd}(\omega )$  is shown in Fig. 8. 
%
\begin{figure}[t]
\includegraphics[width=0.8\columnwidth,height=0.45\columnwidth]{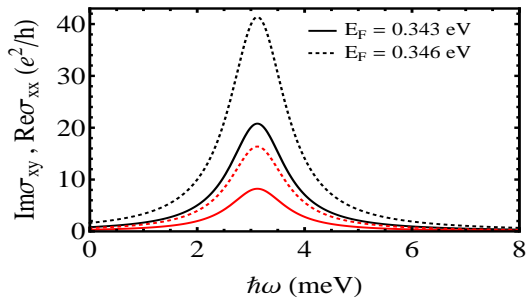}
\vspace*{-0.2cm}
\caption{Intraband limit of the real part (upper two curves) of the longitudinal optical
conductivity and of the imaginary part %\textbf{
(lower two curves) of the Hall conductivity versus %as  functions of the 
photon energy for two values of %the Fermi level 
$E_{F}$ and % field 
$B=10$ Tesla. %as indicated.
The energy 
$\hbar \omega$ is measured from the bottom of the conduction band.}
\end{figure}
%ADD COMMENT ABOUT 
The real part of $\sigma _{xx}^{nd}(\omega )$  is shown in Fig. 8 by the upper two curves. 
As seen, the optical spectral weight under these curves  
 increases with %increasing %the values of 
 $E_{F}$. These peaks lie
in the range of microwave-to-THz frequencies and their amplitude is
larger than that of the interband transitions shown in Fig. 3. This is consistent
with  graphene or topological insulators and other symmetric 2D systems in which  the
%corresponding 
relevant spectral weight increases with %increasing 
$E_{F}$, see, e.g., Fig. 7 of Ref. \cite{PEC},
and the optical  features appear in the THz regime only \cite{CJ,CM,PEC}.
The two lowest curves show  the imaginary part of the Hall conductivity $\sigma _{xy}^{nd}(\omega )$ 
for the same values of $E_{F}$. As seen, apart from the scale, $\sigma _{xy}^{nd}(\omega )$ shows the same behaviour as $\sigma _{xx}^{nd}(\omega )$.
The magnitudes of $\sigma _{xy}^{nd}(\omega )$ and $\sigma _{xx}^{nd}(\omega )$ are different 
due to the different values of the velocity components along the $x$ and $y$ axes.
The peaks of all curves occur at $\omega =\omega _c$.

%ADD COMMENT ABOUT (b), COMBINE (A) AND (b) IN ONE FIG., REMOVE THE SCALE/UNITS
%IN (a) and the parameters in (b).

\section{Oscillator Strength}

The oscillator strength depends strongly on the symmetries of the initial
and final state wave functions and is a function of the in-plane momentum
vector \cite{FMP}. The strength of an optical transition is typically characterized by
the dipole matrix element between the initial and final states. Since a
dimensionless quantity is more useful for making comparisons in different
systems, the oscillator strength is often used instead of the dipole matrix
element. It is defined through the $\mathbf{A\cdot p}$
term in the Hamiltonian describing the interaction between the electron and
the electromagnetic field as
\begin{equation}
O_{n^{\prime },n}={2m\over\hbar^2} \sum_{s,s^{\prime }}(E_{n^{\prime }}^{s^{\prime
}}-E_{n}^{s})  
\left\vert \left\langle n^{\prime },s^{\prime },k_{y}^{\prime }\right\vert 
\mathbf{r}\left\vert n,s,k_{y}\right\rangle \right\vert ^{2},  \label{19}
\end{equation}
where $m=\sqrt{m_{sx}^{\prime }m_{sy}}$ is the mass of the particles 
and  $(E_{n^{\prime }}^{s^{\prime
}}-E_{n}^{s})/\hslash $  the frequency  involved in the  
transitions from the initial to the final states. With the help of Eq. (3),
the relevant matrix element of the position operator 
($\left\langle n^{\prime }\right\vert \mathbf{r}%
\left\vert n\right\rangle \equiv\left\langle n^{\prime },s,k_{y}^{\prime
}\right\vert x\left\vert n,s,k_{y}\right\rangle $) is 
\begin{eqnarray}
\notag
&&\hspace*{-0.7cm}\left\langle n^{\prime },s,k_{y}^{\prime }\right\vert x\left\vert
n,s,k_{y}\right\rangle   \label{20} 
=\Big[ x_{0}\delta _{n^{\prime },n}+\\*
&\hspace*{-0.4cm}+&(1/\sqrt{2}\xi^{s})
\left( 
\sqrt{n+1}\delta _{n^{\prime },n+1}+\sqrt{n}\delta _{n^{\prime },n-1}\right) %
\Big] \delta _{k_{y}^{\prime },k_{y}}.  %\notag
\end{eqnarray}

Equations (19) and (20) clearly show the intraband and interband transitions shown in Figs. 2-8. 
The transitions follow the selection rule $n^{\prime}=n\pm1$.
It is  interesting to physically interpret the oscillator strength 
in terms of intraband and interband transitions.
The intraband transitions between the $n$ and $n+1$ states by the absorption of a photon 
shown in Fig. 8 are the same as those shown by  Eqs. (19) and (20).  For these transitions we have $E_{n+1}^{+}-E_{n}^{+}=\hslash \omega _{c}^{e}, s=s'=+$, whereas the interband 
transitions of Figs. 3-7 follow the rule $n^{\prime}=n\pm1$ but with $s\neq s'$. 
The first absorption peak in the optical longitudinal conductivity
is the sum of the two transitions  $E_{n}^{-}\rightarrow E_{n+1}^{+}$ and 
$E_{n+1}^{-}\rightarrow E_{n}^{+}$. This corresponds to the first two peaks in the optical 
Hall conductivity with the first being positive and the second negative. The corresponding energy absorption for the first peak is
$E_{1}^{+}-E_{0}^{-}$ and for the second one $E_{0}^{+}-E_{1}^{-}$.  Similarly other peaks follow the well-defined  selection rules between higher LLs. The spacing between the peaks also 
depends on the broadening, which we have fixed. When the Fermi level is in the 
band and  not zero,  the peak heights are suppressed and shifted downward.

\section{Summary}

\textit{\ } We studied  magneto-optical transport properties
of monolayer phosphorene subject to an external perpendicular magnetic
field.  The relevant conductivities exhibit periodic oscillations  that can be 
controlled by the magnetic field $B$.  In each band the oscillation peaks are 
equidistant, reflecting the equally
spaced LLs, and show a linear  dependence on $B$. The intraband and interband optical
transitions pertain to two completely different regimes:
the former  occur in the microwave-to-terahertz range and the  latter ones in the visible
frequency range. This is in contrast with a conventional 2D electron gas, 
topological insulators, and graphene in which these features appear only in the 
THz regime. It is also in contrast with phosphorene's responses at B=0 which occur in the mid- to near-infrared regime
 \cite{HL}(a), \cite{MB},  \cite{NYC}.  These  findings expand the horizon of   the optical properties of 2D 
phosphorene and are expected to be useful in the design of new optical devices.

\appendix
\numberwithin{equation}{section}
\section{}

%Using the Hamiltonian given in Eq. (1) of the main text, t
The velocity operator, obtained from Eq. (1), reads
\begin{equation}
v=\frac{\partial H}{\partial\mathbf{p}}=\left(
\begin{array}
[c]{c}%
\alpha^{\prime}p_{x}+\beta p_{y}\\
0
\end{array}%
\begin{array}
[c]{c}%
0\\
-\lambda^{\prime}p_{x}-\eta p_{y}%
\end{array}
\right). \label{A.1}%
\end{equation}
For $s=s'$ the explicit evaluation of the velocity matrix elements gives Eqs. (8) and (9)
with $s=+\equiv e$ and $s=-\equiv h$. For $s\neq s'$ we obtain explicitly 
\begin{eqnarray} 
v_{x,n,n^{\prime}}^{-,+} &  =&  
i\, (v_{x}^e- v_{x}^h) \,\delta_{k_{y},k_{y}^{\prime}} 
{\displaystyle\int\limits_{-\infty}^{\infty}}
%EndExpansion
dx\,\phi_{n}(u^{h})\nonumber\\*
&\times &\hspace*{-0.2cm}\,\,\Big[\sqrt{n^{\prime}+1}\phi_{n^{\prime}+1}(u%
^{e})-\sqrt{n^{\prime}}\phi_{n^{\prime}-1}(u^{e})\Big].\,  
\end{eqnarray}
Because $u^{h}=1.1u^{e}$ we set $u^{h}=u^{e}$ in order to have 
simple expressions. We then have  
\begin{equation}% 
I={\displaystyle\int%\limits_{-\infty}^{\infty}
} 
dx\,\phi_{m}^{\ast}(u^{h})\phi_{n}(u^{e})\approx\delta_{n,m}\text{.}\label{A.10}%
\end{equation}
This gives $v_{x,n,n^{\prime}}^{-,+}=v_{x,n,n^{\prime}}^{+,-}$ and $v_{y,n^{\prime},n}^{-,+}=v_{y,n^{\prime},n}^{+,-}$  with  
\begin{eqnarray}%{equation}
\hspace*{-0.8cm}v_{x,n,n^{\prime}}^{-,+}&=&i\{v_{x}^e
-v_{x}^h\}\Big[ \sqrt{n}%^{\prime}+1}%
\delta_{n,n^{\prime}+1}-\sqrt{n^{\prime}}\delta_{n,n^{\prime}-1}\Big]\delta_{k_{y},k_{y}^{\prime}}, 
\label{A.11}\\*%
%\end{equation}
%
%and
% 
%\begin{equation}
\hspace*{-0.8cm}
v_{y,n^{\prime},n}^{-,+}&=&\{v_{y}^e
-v_{y}^h\}\Big[\sqrt{n^{\prime}}%+1}
\delta_{n^{\prime},n+1}+\sqrt{n}\delta_{n^{\prime
},n-1}\Big] \delta_{k_{y},k_{y}^{\prime}}.
\label{A.12}%
\end{eqnarray}%{equation} 

%multiplied by $\delta_{k_{y},k_{y}^{\prime}}$.
To check the approximation (A.3) we evaluate explicitly the integral $I$ for $n=m$
($I$ vanishes for $n\neq m$, see Ref. \cite{gr}) and $n=0,1,2, 5, 10$ using the explicit  expressions of the Hermite polynomials, e.g.,  $H_{0}(x)=1,H_{1}(x)=2x,H_{2}(x)=4x^{2}-2$, etc. 
The  values we obtain for $n=0, 1, 2, 5,10$ are, respectively,
$0.997, 0.992, 0.983, 0.924$,  and $0.742$. 
This shows that the approximation (A.3) is a valid one at least when the magnetic field is strong and only a few LLs are occupied. \\

\vspace*{-0.2cm}
{\bf Acknowledgments}:
This work was supported by the the Canadian NSERC Grant No. OGP0121756 (MT, PV) and by the Flemish Science Foundation (FWO-Vl) (FMP).\\

\noindent Electronic addresses: \\$^{\star}$m.tahir06@alumni.imperial.ac.uk,
$^{\dag}$p.vasilopoulos@concordia.ca, $^{\ddag}$francois.peeters@uantwerpen.be

\end{document}